\documentclass[3p,a4paper,times,twocolumn]{elsarticle}

\usepackage{graphics}
\usepackage{lineno}
\usepackage{subcaption}	

\usepackage{amsmath,amssymb,amsfonts,amsthm,amsbsy,amstext,stmaryrd}
\usepackage{siunitx}		
\usepackage{xfrac}

\usepackage{booktabs} 	
\usepackage{multirow} 	

\usepackage{color}

\usepackage{url}

\newcommand{\etal}{\emph{et al.}\ }

\biboptions{square,comma,sort&compress}

\journal{Acta Materialia}

\begin{document}

\begin{frontmatter}

\title{From solid solution to cluster formation of Fe and Cr in $\alpha$-Zr}

\author[IC,ANSTO]{P.A.~Burr\corref{corresponding}}\ead{burr.patrick@gmail.com}
\author[IC]{M.R.~Wenman}
\author[Oxford]{B.~Gault}
\author[Oxford]{M.P.~Moody}
\author[Sandvik,Manchester]{M.~Ivermark}
\author[IC]{M.J.D.~Rushton}
\author[Manchester]{M.~Preuss}
\author[ANSTO]{L.~Edwards}
\author[IC]{R.W.~Grimes}
\address[IC]{Centre for Nuclear Engineering and Department of Materials, Imperial College London, London, SW7 2AZ, UK.}
\address[ANSTO]{Institute of Materials Engineering, Australian Nuclear Science \& Technology Organisation, Menai, New South Wales 2234, Australia.}
\address[Oxford]{Department of Materials, University of Oxford, Parks Road, Oxford, OX1 3PH, UK.}
\address[Sandvik]{High Temperature Materials, Sandvik Materials Technology, 734 27 Hallstahammar, Sweden.}
\address[Manchester]{University of Manchester, School of Materials, M13 9PL, UK.}
\cortext[corresponding]{Corresponding author.}

\begin{abstract}
To understand the mechanisms by which the re-solution of Fe and Cr additions increase the corrosion rate of irradiated Zr alloys,
the solubility and clustering of Fe and Cr in model binary Zr alloys was investigated using a combination of experimental and modelling techniques --- atom probe tomography (APT), x-ray diffraction (XRD), thermoelectric power (TEP) and density functional theory (DFT). 
Cr occupies both interstitial and substitutional sites in the $\alpha$-Zr lattice; Fe favours interstitial sites, and a low-symmetry site that was not previously modelled is found to be the most favourable for Fe. Lattice expansion as a function of Fe and Cr content in the $\alpha$-Zr matrix deviates from Vegard's law and is strongly anisotropic for Fe additions, expanding the $c$-axis while contracting the $a$-axis. Matrix content of solutes cannot be reliably estimated from lattice parameter measurements, instead a combination of TEP and APT was employed.
Defect clusters form at higher solution concentrations, which induce a smaller lattice strain compared to the dilute defects.
In the presence of a Zr vacancy, all two-atom clusters are more soluble than individual point defects and as many as four Fe or three Cr atoms could be accommodated in a single Zr vacancy. The Zr vacancy is critical for the increased apparent solubility of defect clusters; the implications for irradiation induced microstructure changes in Zr alloys are discussed. 
\end{abstract}

\end{frontmatter}



\section{Introduction}
\label{sec:intro}

Zr alloys are widely used in the nuclear industry as nuclear fuel cladding and other structural components. Fe and Cr are common alloying elements, added to improve corrosion resistance \cite{Garzarolli1996,Sabol2005,Garzarolli2011}. These elements are known to exhibit near-negligible solid solubility in $\alpha$-Zr, and therefore segregate to form second phase particles (SPPs) \cite{Chemelle1983,Barberis2005,Lumley2013}.
One of the key aspects of the microstructural evolution of Zr alloys under irradiation is the amorphisation and subsequent dissolution of SPPs, which leads to the re-solution of the alloying elements in the Zr matrix above their solubility limits \cite{Yang1986,Yang1988,Griffiths1987,Griffiths1987a,Griffiths1988,Lefebvre1990,Griffiths1990,Griffiths1995,Kakiuchi2006,Griffiths2011,Cockeram2011,Cockeram2014,Francis2014}.
In turn, this has an impact on the physical and corrosion properties of the alloy. In particular, surface oxidation and hydrogen pick-up fraction are known to be strongly affected by alloy composition and the presence and distribution of SPPs and experience a marked increase after SPP dissolution \cite{Pecheur1992,Barberis2002,Tagtstrom2002,Kakiuchi2004,Strasser2008,Valizadeh2010,Garzarolli1996,Garzarolli2011,Burr2012}. It is important to limit hydrogen uptake during reactor operation because hydrogen causes dimensional changes to the cladding \cite{Strasser2008}, reduces its ductility \cite{Strasser2008} and reduces integrity performance in hypothetical accident scenarios \cite{Strasser2008a,Raynaud2011}, and potentially in the storage conditions relevant to spent nuclear fuel \cite{Strasser2008a,Kessler2002}.


Recent advanced transmission electron microscopy (TEM) \cite{Harte2015a} and atom probe tomography (APT) studies \cite{Sundell2014} have shown that clusters of Fe and Cr form at $\left< a \right>$ and $\left< c \right>$ dislocation loops following the re-solution process. This was previously suggested by TEM investigation \cite{Griffiths1987,Griffiths1987a,Griffiths1988,Cockeram2011,Cockeram2013,Cockeram2014,Cockeram2015} but not observed directly.
It has been suggested that irradiation induced defects may also act as trapping sites for hydrogen, thereby increasing the terminal solid solubility of hydrogen in $\alpha$-Zr \cite{McMinn2000,Vizcano2002}.

The solubility of Fe in Zr --- and to a lesser extent also that of Cr in Zr --- has also been investigated using atomic scale simulations, but so far, the clustering behaviour of alloying elements has hardly been considered using such methods.
Early work by Per\`ez and Weissmann \cite{Perez2008} and of Pasianot \etal \cite{Pasianot2009} investigated the possible mechanism for Fe accommodation in the $\alpha$-Zr lattice, but their DFT calculations were limited to small supercells containing 36 and 48 Zr atoms respectively. In particular, Pasianot \etal \cite{Pasianot2009} observed that when Fe substitutes for Zr, it occupies a low symmetry configuration that is displaced slightly from the lattice site.
More recent studies have employed, in one case, slightly larger supercells (54 Zr atoms by Lumley \etal\cite{Lumley2013} and 48 Zr atoms by Christensen \etal\cite{Christensen2014,Mader2013}), but only the more conventional interstitial sites (tetrahedral and octahedral) were considered. Furthermore, recent publications \cite{Willaime2003,Pasianot2012,Samolyuk2013,Peng2012,Peng2013,Varvenne2013,Verite2013} have shown that even larger supercells ($\sim300$ atoms if no finite size correction term is applied) are required to avoid computational artefacts that may significantly affect the apparent stability of defects in $\alpha$-Zr.
It is evident that a state-of-the-art evaluation of the extrinsic defects in Zr is needed.

In previous papers, the authors considered the formation of SPPs \cite{Lumley2013,Ivermark2011} and their interaction with H \cite{Burr2012,Burr2013}.
Here, the authors are concerned with the conditions relevant to irradiated Zr alloys, in which the SPPs are partially dissolved.
The current work employs a suite of experimental and theoretical approaches to investigate the solubility of Fe and Cr in pristine and defective Zr and the formation of clusters containing Fe, Cr and intrinsic defects. 
First, DFT simulations reveal that  Cr may occupy both interstitial and substitutional sites in the $\alpha$-Zr lattice, and the results were corroborated by spatial distribution maps produced with APT. The simulations also indicated that the Fe-Zr binary systems exhibits a large deviation from Vegard's law, thereby indicating that lattice parameter measurements by XRD do not provide a suitable estimate of solute concentration in the $\alpha$-Zr matrix. The matrix content of fast quenched binary samples was then measured using TEP and APT, showing that an increasing amount of Fe and Cr was trapped in solution with increasing nominal composition, despite the formation of SPPs.
The lattice expansion due to alloying additions of the binary samples was then measured by XRD, and the predicted deviation from Vegard's law was observed. Complementary DFT simulations also highlight that the preference for interstitial over substitutional accommodation is dependent on the atomic strain environment and an argument is put forward for clustering of dilute defects as a means to reduce overall lattice strain.
A first nearest neighbour analysis carried out with APT provides evidence that these clustering tendencies occur in oversaturated Cr-Zr alloys.
Further simulations show that larger defect clusters may be favourably accommodated in the $\alpha$-Zr lattice and that Zr vacancies are crucial for the formation and growth of the clusters.
Finally, the work is summarised and the implications for irradiated Zr alloys discussed. 

\section{Methodology}								\label{sec:meth}
\subsection{Materials}								\label{sec:methMater}

Binary Zr-alloys were melted in an arc furnace in a water cooled copper crucible under an argon atmosphere at Western Zirconium, USA. The Zr starting material was in the form of chips while Cr and Fe were small beads. All alloying elements were standard materials used by Western Zirconium for their production of zirconium alloys. The 125 g buttons were re-melted three times in order to ensure chemical homogeneity.
Further details of the materials processing can be found in \cite{Gault2013,Ivermark2009}. The buttons were analysed at Western Zirconium using induced coupled plasma-atomic emission spectroscopy and combustion analysis (oxygen and nitrogen) to determine the chemical constituent of each sample, presented in Table~\ref{tab:ICP}.
It is acknowledged that sample \emph{Zr-0.05Cr} contains notable amounts of Fe and Sn contaminations and the results from this alloy are highlighted in subsequent section. All other alloys were produced with a high degree of purity.

\begin{table}[hb]
\centering
\small
\caption{\label{tab:ICP} Chemical composition of the binary alloys in wt.~ppm. Hf and Nb were consistently less then 23 and 20 ppm respectively. Si and any other potential impurities were always below the detection limit.}
\begin{tabular}{l r r r r r r r r}
\toprule
Sample name 	&\multicolumn{1}{c}{Cr}&\multicolumn{1}{c}{Fe}	&\multicolumn{1}{c}{Cu}&\multicolumn{1}{c}{N}&\multicolumn{1}{c}{O}&\multicolumn{1}{c}{Sn} \\
\midrule
Zr-0.1Fe		&$<20	$&$1049	$&$10	$&NA	  &NA	  &$<8	$\\
Zr-0.2Fe		&$<20	$&$1927	$&$11	$&NA	  &NA	  &$<8	$\\
Zr-0.4Fe		&$<20	$&$4298	$&$<10	$&$44	$&$810	$&$<8	$\\
Zr-0.6Fe		&$<20	$&$6226	$&$22	$&NA	  &NA	  &$<8	$\\
Zr-0.8Fe		&$<20	$&$8943	$&$19	$&NA	  &NA	  &$<8	$\\
\midrule
Zr-0.05Cr		&$475	$&$217	$&$11	$&NA	  &NA	  &$1155	$\\
Zr-0.15Cr		&$1608	$&$37	$&$<10	$&NA	  &NA	  &$<8	$\\
Zr-0.30Cr		&$2869	$&$41	$&$<10	$&$43	$&$849	$&$<8	$\\
\bottomrule
\end{tabular}
\end{table}

The as-cast buttons were cross rolled at \SI{540}{\degree C} with an intermediate recrystallisation anneal at \SI{600}{\degree C} to a final thickness of 3 mm. Subsequently, $3 \times 3 \times 40$ \si{mm^3} matchstick samples were cut and $\beta$ heat-treated for 10 minutes at \SI{1000}{\degree C} in a vertical furnace flushed with argon, followed by water quenching, in an attempt to maintain most of the Fe and Cr into $\alpha$-Zr solution. Scanning and transmission electron microscopy investigation showed that complete solid solutions were not obtained even at these very high cooling rates. Instead a significant number of small SPPs had formed \cite{Ivermark2009}.

\subsection{X-ray diffraction}									\label{sec:methXRD}
The matchsticks were cut into $3 \times 3 \times 2$~\si{mm^3} cubes, mounted in a $5 \times 5$ grid before grinding and polishing to produce an XRD-sample with an approximate surface dimension of $15 \times 15$~\si{mm^2}. The x-ray analysis was carried out on a Philips PW3710 diffractometer using Cu-K$_{\alpha}$ radiation in Bragg-Brentano geometry. Diffraction profiles were recorded ranging from the $\{10\bar{1}0\}$ to the $\{30\bar{3}2\}$ reflections with a step size of $2\theta = 0.02$\si{\degree} and a recording time of \SI{20}{s} per step. The profiles were analysed using Rietveld analysis to determine the $a$ and $c$ lattice parameters.

\subsection{Thermoelectric power measurements}					\label{sec:methTEP}
TEP experiments measure the Seebeck coefficient (S), which is the electric potential difference that arises when two metals in tight contact form a thermocouple with two junctions held under a temperature difference. The Seebeck effect consists of two parts: a chemical gradient found at the junctions between the two metals (the Peltier effect) and a thermal gradient within the same metal (the Thomson effect). In the present case, matchstick samples were clamped between copper blocks, which were maintained at temperature T and $\text{T}+\Delta\text{T}$, respectively. The measured TEP, which is relative to the reference metal to which it is clamped, can then be plotted as a function of concentration of solid solution in the matrix \cite{Pelletier1978,Borrelly1978,Pelletier1982,Pelletier1984}. 
The sign of the TEP can be discussed in terms of incomplete d-bands of electrons in transition metals \cite{Potter} and Fermi surfaces \cite{Bruno1994}. For an element with unfilled d-orbitals, such as Zr, the addition of a solute atom into the matrix can either increase or decrease the TEP of the alloy. As all transition metals have fewer available d-orbitals than Zr, the energy difference decreases, which results in the TEP becoming more negative with increasing solute concentrations \cite{JinKim2002}.
Previous work has shown that the Seebeck coefficient of zirconium alloys is sensitive to solute concentration, texture and cold work, but is not affected by the presence of small volume fraction of SPPs \cite{Merle1986,Borrelly1990,Loucif1992,Loucif1993}.
All samples were prepared using the same procedure to minimise variations in texture and microstructure (see section~\ref{sec:methMater}), therefore any change recorded in TEP can be attributed to variations in solute concentrations.

The measurements were conducted at INSA, Lyon, France using a TechMetal Promotion instrument and a Cu reference. The temperature of the clamping copper blocks was held at $15 \pm 0.2$ \si{\degree C} and $25 \pm 0.2$ \si{\degree C}. In order to stabilise the thermo-electricity, each specimen was left for \SI{1}{min} after mounting before measurement. Each measurement had a duration of \SI{20}{s} in which the initial value and the variation from this value were recorded. Each surface of the matchstick specimen was measured twice to give an average value from eight measurements per alloy concentration.

\subsection{Atom probe tomography}							\label{sec:methAtom}
\begin{figure*}[hbt]
\centering
\includegraphics[width=\textwidth]{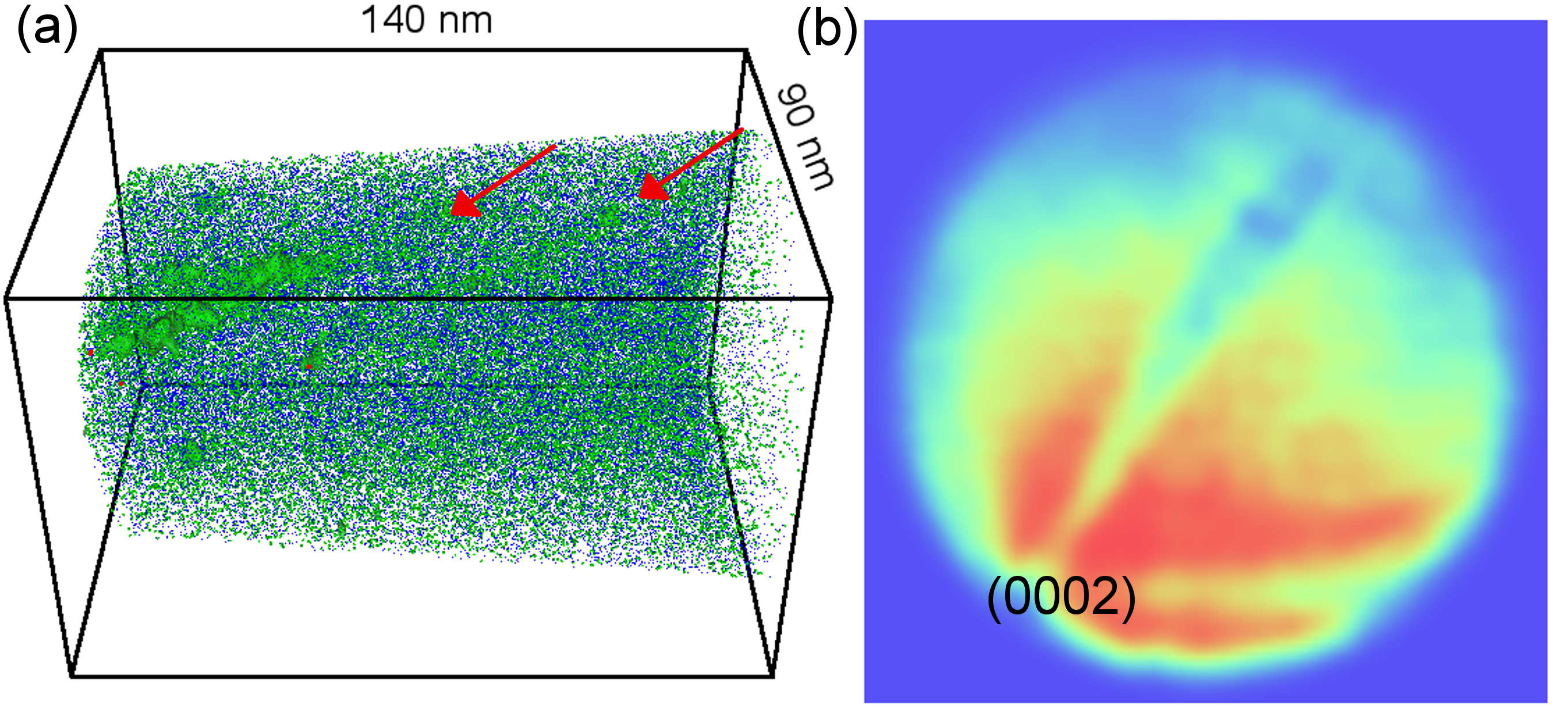}
\caption{\label{fig:APT} (a) Three-dimensional reconstruction of the dataset from the Zr-Cr sample. The green surface encompasses regions containing above 2~at.\% Cr. (b) Top-down projection of the dataset shown in (a), with red representing regions of higher density and blue of lower density.}
\end{figure*}

APT was performed on a Cameca LEAP 3000 X Si, with a flight path of \SI{90}{mm}. The experiments were conducted at a base temperature of $-213\pm5$\si{\celsius}, in laser-pulsing mode (\SI{\sim10}{ps}, \SI{532}{nm}, spot size \SI{< 10}{\micro m} diameter). Throughout the analysis, the DC voltage was increased to keep a detection rate of 5 ions per 1000 pulses. Specimens were prepared by means of a FIB lift-out procedure, from a mechanically polished sample of the Zr-alloys, using a Zeiss Auriga and an electropolished Mo grid as support \cite{Felfer2012}. The datasets were reconstructed using state-of-the-art algorithms \cite{Gault2011}, resulting in the typical tomograms shown in Figure~\ref{fig:APT}(a) for a binary Cr-Zr specimen containing \SI{0.26}{at.\% Cr}. Therein, a \SI{2}{at\% Cr} isoconcentration surface highlights the presence of small regions enriched in Cr, up to 4--5 at\%, which appear to be aligned, maybe along a twin boundary, similar to what was discussed in \cite{Gault2013}.  
The signal to background noise ratio on the major peak of Cr was 120:1, and on the major peak of Fe it was above 200:1, providing a high level of certainty when labelling the Cr and Fe atoms. 
Isotopic ratio of $^{56}$Fe/$^{54}$Fe was found to be 14.3, which compares well with the natural abundance (15.81). This provides confidence that the signal was mostly generated from Fe ions and not molecular impurities such as CO. 

As evidenced in Figure~\ref{fig:APT}(b), which contains a two-dimensional density map computed from the same tomogram, the data exhibits features that can be directly related to the crystallography of the specimen \cite{Gault2012}: a so-called pole can be seen at the bottom left along with a series of zone lines exhibiting six-fold symmetry. This pole can be attributed to the (0002) atomic planes at the specimen surface causing trajectory aberrations. These planes are also imaged within the APT data. Imaging atomic planes is common in APT, and allows for calibration of the tomogram  \cite{Gault2009,Gault2008} as well as site occupancy analyses \cite{Boll2007,Miller2011,Gault2012a,Sakurai1992}, which are facilitated by data treatment methods such as spatial distribution maps \cite{Boll2007,Geiser2007,Moody2009}. The latter are similar to split radial distribution functions investigating the local neighbourhood of each atom along a specific direction. Another common data treatment technique, the calculation and analysis of the distance between a given species and its first nearest neighbours \cite{Shariq2007,Stephenson2007,DeGeuser2011}, was also used here to estimate the matrix composition following the protocol described in \cite{DeGeuser2011} as reported in \cite{Gault2013}, but also to investigate the clustering tendency of Cr in Zr.

\subsection{Computational calculations}					\label{sec:methDFT}
DFT simulations were carried out using the {\sc castep} code \cite{Clark2005} with the PBE exchange-correlation functional \cite{Perdew1996}, ultra-soft pseudo potentials \cite{Vanderbilt1990} and a consistent plane-wave cut-off energy of \SI{450}{eV}.


Supercells containing 150 Zr atoms were modelled using a $2\times2\times2$ {\bf k}-point sampling grid \cite{Monkhorst1976}. The linear elastic theory correction term of Varvenne \etal \cite{Varvenne2013} (\emph{aneto}) was employed to reduce finite size effects.
The elastic constants of $\alpha$-Zr, which feed into \emph{aneto}, were calculated by performing small lattice perturbations. The resulting stiffness constants are (in units of GPa): $c_{11} = 141.97$, $c_{12} = 65.36$, $c_{13} = 68.02$, $c_{33} = 148.71$, $c_{44} = 30.22$ and $c_{66} = 38.30$.
Since these systems are metallic, density mixing and Methfessel-Paxton \cite{Methfessel1989} cold smearing of bands were used (smearing width $ = 0.1$~eV). Tests were carried out to ensure a convergence of \SI{1e-3}{eV/atom} with respect to all parameters.
No symmetry operations were enforced when calculating point defects and all simulations were spin polarised.

The energy convergence criterion for self-consistent calculations was set to \SI{1e-8}{\electronvolt}. 
Similarly robust criteria were imposed for atomic relaxation within the memory conservative BFGS algorithm \cite{Byrd1994,Pfrommer1997}: the energy difference was less than \SI{1e-6}{\electronvolt}, forces on individual atoms less than \SI{0.1}{\electronvolt\per\nano\meter} and for constant pressure calculations, stress components on the cell of less than \SI{1}{\mega\pascal}.

Defect formation energies $E^f$ were calculated using equation \ref{eq:def_form}.
\begin{equation}\label{eq:def_form}
E^f = E^{\text{DFT}}_d - E^{\text{DFT}}_p \pm \sum_i \mu(i) + \frac{1}{2} E_{\text{int}}
\end{equation}
where $E^{\text{DFT}}_d$ and $E^{\text{DFT}}_p$ are the total energies from the defective and perfect DFT simulations, $\mu_{i}$ is the chemical potential of all species $i$ that are added or removed from the perfect crystal to form the defect, and $E_{\text{int}}$ is the correction term for the interaction energy of the defect with its periodic images, calculated using \emph{aneto} \cite{Varvenne2013}.
The chemical potential $\mu$ is calculated as the DFT energy per atom of the metallic elements in their ground state; for Fe the ground state is the ferromagnetic BCC phase, for Cr it is the anti-ferromagnetic BCC phase.
\emph{Aneto} calculations employed a radial cutoff of \SI{15}{\angstrom} with 20 divisions of the Fourier grid, which yielded energy values converged up to the $4^{\text{th}}$ decimal place.

The relaxation volume ($\Omega$) of a defect is defined as the difference in volume between a supercell containing the defect and the perfect Zr supercell; see equation~\ref{eq:relVol}.
\begin{equation}
\label{eq:relVol}
\Omega = V(Zr_xM_y) - V(Zr_z)
\end{equation}
When calculating $\Omega$, mass action is not taken into account, in other words, the number and types of atoms between the defective and perfect cell do not have to be the same. In fact, considering the subscript of equation~\ref{eq:relVol}, for an interstitial defect $x=z$, whilst for a substitutional defect $x+y=z$.
A related quantity often found in the literature is the defect formation volume, in which the number of atoms of each species is conserved between perfect and defective cells. However, the defect formation volume is only properly defined for intrinsic defects \cite{Bruneval2012}, as the reference volume of an isolated extrinsic atom is not a strictly defined quantity. In the current work, only relaxation volumes will be considered.

Configurational averaging of physical properties related to the presence of defects (such as $\Omega$) was performed following equation~\ref{eq:configAverage}:
\begin{equation}
\label{eq:configAverage}
 \bar{X} = \frac{\sum_i{n(i) X(i) \exp(-Q_i)}}{\sum_i{n(i) \exp(-Q_i)}}
\end{equation}
where
\begin{equation}
 Q_i = \frac{\Delta E_f(i)}{k_BT}
\end{equation}
and $X$ is the physical property of interest, $i$ is a defect configuration, $n(i)$ is its multiplicity, $\Delta E_f(i)$ is the difference in formation energy with respect to the most stable defect and all other symbols retain their conventional meaning.


\section{Results and Discussion}
\subsection{Fe-Zr and Cr-Zr binary series}								\label{sec:makingbuttons}

\begin{figure}[hbt]
\centering
\includegraphics[width=0.49\textwidth]{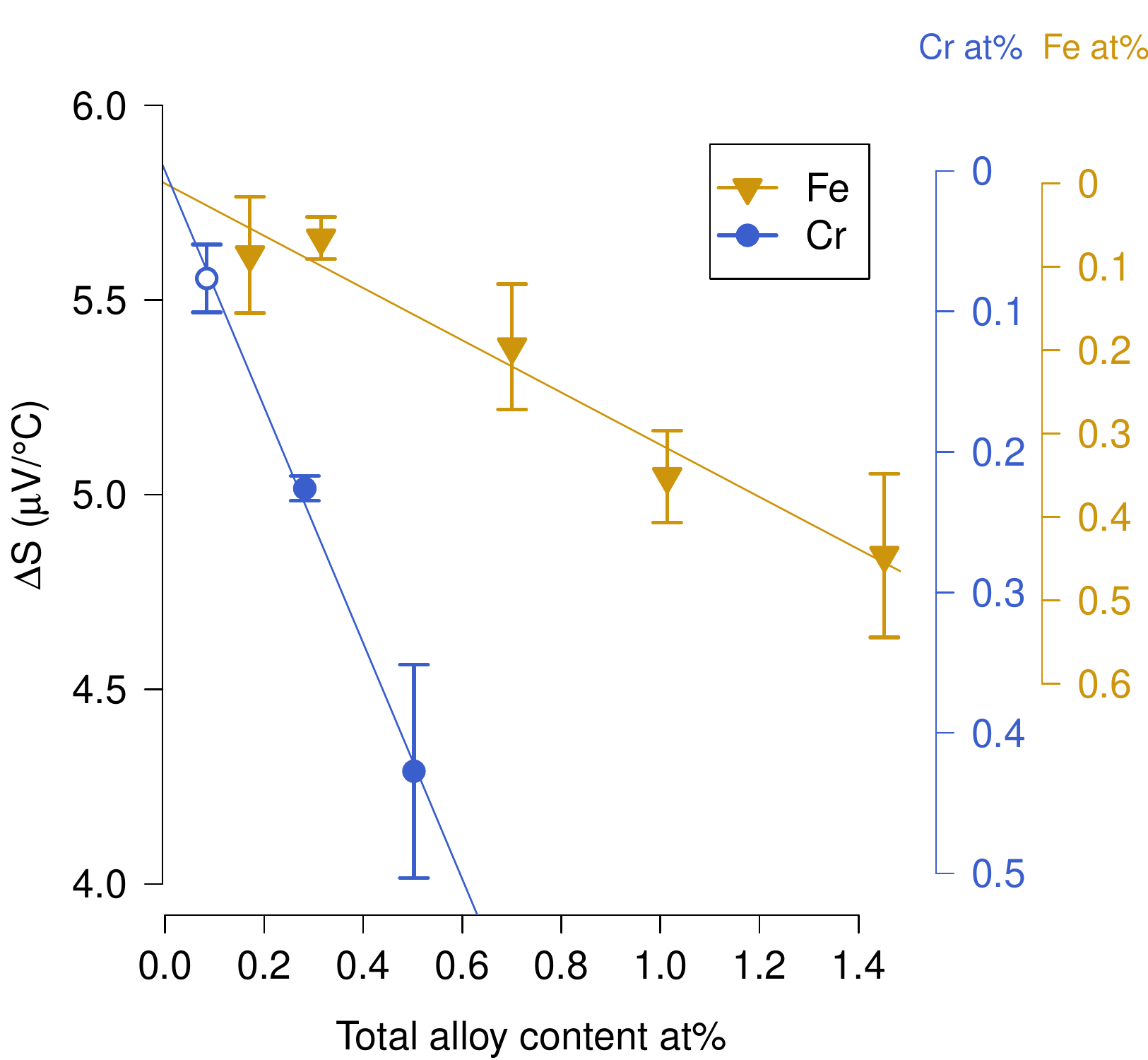}
\caption{\label{fig:TEP} TEP measurements against nominal composition of the binary alloys. Right-hand side axes are the calibrated concentrations of Cr and Fe dissolved in solution. Error bars express the standard deviation of eight samples per datum. The hollow symbol represent the low purity alloy.}
\end{figure}

TEP measurements of the samples evaluate the Seebeck coefficient of the $\alpha$-Zr matrix, bypassing the resistance contributions of the SPPs \cite{Pelletier1978,Borrelly1978,Pelletier1982,Pelletier1984,Borrelly1990,Loucif1992,Loucif1993}. Therefore, any change in Seebeck coefficient observed across the binary alloys, compared to the unalloyed reference sample, can be related to the amount of Fe or Cr in solution. TEP results are plotted against composition in Figure~\ref{fig:TEP}. It is observed that the Seebeck coefficient decreases with increasing content of Fe or Cr, indicating that, despite the formation of SPPs, an increasing amount of alloying element was trapped in solution with increasing nominal composition of the samples.

To obtain quantitative concentrations for the $\alpha$-Zr matrix, it is necessary to establish a datum by calibrating at least one TEP point that has a known alloy concentration in solution. Since all of the samples exhibited some segregation of Fe/Cr to SPPs, APT was employed to calculate the matrix content of Fe and Cr in $\alpha$-Zr by sampling a volume containing no SPPs. The matrix content, therefore, includes all the Fe and Cr atoms that are not within SPPs, irrespective of the atoms being uniformly dispersed in solution, segregated at the grain boundaries or clustered in atmospheres of higher Cr or Fe density. These results were published previously~\cite{Gault2013}, and the relevant findings are summarised thus.
The samples containing 1.30~at.\% Fe and 0.26~at.\% Cr were chosen for APT analysis. 
Fe and Cr are either found within small particles or atmospheres along grain boundaries, or are randomly distributed within the matrix.
The total composition calculated using APT is in excellent agreement with the nominal composition of the alloys ($1.34\pm0.026$~at.\%~Fe and $0.25\pm0.01$~at.\%~Cr respectively). Selecting volumes that contained no SPPs or grain boundaries, the concentration of alloying element in solution was calculated to be $0.42\pm0.015$~at.\%~Fe and $0.21\pm0.01$~at.\%~Cr.

The change in Seebeck coefficient measured by TEP is reasonably linear with increased alloy concentration (see Figure~\ref{fig:TEP}).
Therefore, assigning the concentration values obtained by APT to the points at 1.30~at.\% Fe and 0.26~at.\% Cr, and extrapolating linearly so that the reference sample has an alloy concentration of zero, the solution concentration of the remaining samples was obtained. The resulting scales are plotted on the right-hand y-axis in Figure~\ref{fig:TEP}.
The deviation from the linear regression then represents a new estimate of the uncertainty, which is added to the uncertainty arising from ATP measurements and that of TEP measurements, to form the total error about the solution concentration of each sample (used in later graphs).
In the following sections, when referring to alloy composition, we refer to these calculated concentrations of Cr or Fe in solution rather than the nominal composition of the buttons.

\subsection{Fe and Cr solubility in $\alpha$-Zr}								\label{sec:interstitials}
To calculate the solution energy of the alloying element within the non-interacting regime, DFT simulations were performed with single Fe and Cr defects in the supercells described previously, equivalent to an alloy concentration of \SI{0.67}{at.\%} Fe/Cr.
A number of recent studies \cite{Willaime2003,Pasianot2012,Samolyuk2013,Peng2012,Peng2013,Varvenne2013,Verite2013} have highlighted that finite cell size effects may significantly affect the predicted stability of Zr self interstitials atoms (SIAs); simulations performed using a small size of supercell, or without the use of finite-size correction methods, yield spurious results.
In particular, the work by Varvenne \etal \cite{Varvenne2013} showed that with an energy correction term calculated from linear elastic theory, as used in the current work ($E_{int}$), supercells containing 200 Zr atoms accurately described Zr SIAs, and supercells containing only 96 Zr atoms yielded differences of only 40--150~meV.
All the defects considered in the current work cause significantly less lattice strain than Zr SIAs. This provides confidence that the supercell employed in this study, which contained 150 Zr atoms, is sufficiently large to avoid spurious finite size effects.
This is corroborated by the fact that $E^p_{int}$ was in the range of 3--60 meV for all point defects.
As a further confirmation, all point defect simulations were repeated under constant volume conditions ($\varepsilon =0$) and constant pressure conditions ($\sigma = 0$), and the difference in energy between the two methods was consistently less then \SI{0.6}{\%}.

Many interstitial positions, as well as the substitutional and Zr-Fe and Zr-Cr dumbbell configurations were considered; 
The resulting formation energies ($E_f^{\varepsilon=0}$), relaxation volumes ($\Omega^{\varepsilon=0}$) and anisotropic strains on the supercell ($\varepsilon_{11}^{\sigma =0}$, $\varepsilon_{33}^{\sigma=0}$) are reported in Table~\ref{tab:interst} for all stable defects.\footnote{$E_f^{\sigma=0}$ and $\Omega^{\sigma=0}$ were within $0.07 \%$ and $3.7 \%$ of $E_f^{\varepsilon=0}$ and $\Omega^{\varepsilon=0}$ respectively.}
Some interstitial positions were found to be unstable, that is, the defects moved to another site upon relaxation; these include the tetrahedral positions (which appears as stable when simulated in small supercells), the hexahedral position (also termed basal tetrahedral by some authors) and all the dumbbell configurations.
The substitutional defect consistently relaxed to the off-site substitutional position discussed in \cite{Perez2008} unless symmetry constraints were imposed. This suggests that the high spin high-symmetry substitutional site observed by Christensen \etal \cite{Christensen2014} may be due to insufficiently high degree of convergence during geometry relaxation.

Similarly, the tetrahedral site relaxed into the newly observed crowdion configuration if a suitably large simulation cell is employed.
The current work identifies another interstitial site that has not previously been simulated: the off-site octahedral. This is significantly more stable than any other interstitial site for Fe, but is unstable for the accommodation of Cr, which is consistent with the larger atomic radius of Cr. M\"ossbauer studies of Fe in $\alpha$-Zr \cite{Yoshida1988} suggested that $\sim30\%$ of the total Fe in solution is located in off-centre interstitial sites of this type.


\begin{table*}[hbt]
\small
\centering
\caption{\label{tab:interst} Defect formation energy ($E_f^{\varepsilon=0}$) and volumetric properties for all defects that may accommodate Fe or Cr in bulk $\alpha$-Zr.
$\varepsilon_{11}^{\sigma=0}$ and $\varepsilon_{33}^{\sigma=0}$ are the strain in the $a$ and $c$ direction respectively. Full relaxation volume tensors are presented in~\ref{App:relaxTensor}.}
\begin{tabular}{l l S S S S}
\toprule
&			&\text{$E_{f}^{\varepsilon=0}$ (eV)}	&\text{$\Omega^{\varepsilon=0}$ (\si{\angstrom^3})}	&\text{$\varepsilon_{11}^{\sigma=0}$ $(\%)$}	&\text{$\varepsilon_{33}^{\sigma=0}$ $(\%)$} \\
\midrule
Fe	&off-site substitution		&1.388	&-10.40	&-0.15	&-0.15	\\
	&substitution			&1.709	&-16.25	&-0.02	&-0.40	\\
 	&octahedral			&1.079	&13.70	&-0.02	&0.27	\\
	&off-site octahedral 		&0.941	&13.53	&-0.07	&0.25	\\
	&trigonal				&1.212	&13.42	&0.18	&-0.26	\\
	&crowdion				&1.172	&13.52	&0.10	&0.09	\\
\midrule
Cr	&off-site substitution		&1.732	&-11.31	&-0.03	&-0.21	\\
	&substitution			&1.892	&-12.83	&-0.11	&-0.13	\\
 	&octahedral			&1.882	&15.20	&-0.05	&0.30	\\
	&trigonal				&1.968	&13.65	&0.16	&-0.17	\\
	&crowdion				&2.061	&15.41	&0.15	&0.08	\\

\bottomrule
\end{tabular}
\end{table*}

Regarding the accommodation of Fe, all interstitial sites provide more favourable solution energies than the substitutional site. This is in agreement with experimental diffusivity measurements~\cite{Hood1972,Hood1988,Perez2003}, and all previous DFT calculations \cite{Perez2008,Pasianot2009,Lumley2013} with the sole exception of \cite{Christensen2014}.
It is unclear from the literature whether Cr atoms exhibit a similar preference for interstitial accommodation.
 Experimental diffusivity measurements indicated that Cr diffuses 2--4 orders of magnitude slower than Fe in $\alpha$-Zr. However, Fe is reported to diffuse 6--9 orders of magnitude faster than for Zr self-diffusion and 9--16 orders of magnitude faster than substitutional solutes \cite{Balart1983,Tendler1987,Hood1988,Perez2003}. This suggests that the transport of Cr in the $\alpha$-Zr lattice may be mediated by an interstitial solute.
The current work is in excellent agreement with a previous DFT publication \cite{Lumley2013}, which highlights that whilst the preferred site for Cr solution is substitutional, the difference in energy between that and the interstitial octahedral site is very small, and therefore Cr is expected to exhibit both substitutional and interstitial behaviour.
Despite this, subsequent modelling studies have not considered the possibility of interstitial accommodation for Cr \cite{Christensen2014}.

\begin{figure}[hbt]
\centering
\includegraphics[width=0.49\textwidth]{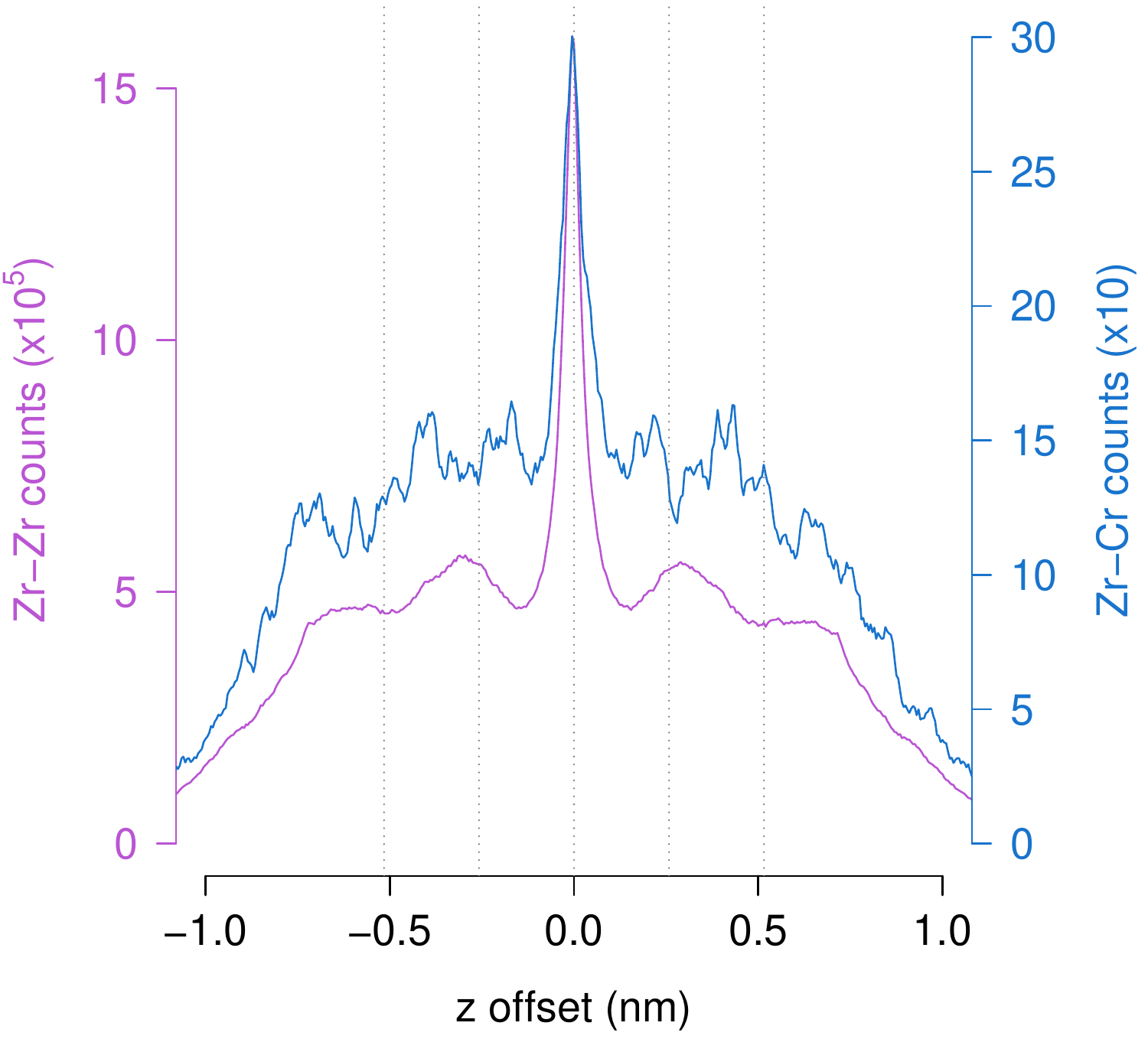}
\caption{\label{fig:proxigram} In-depth spatial distribution map showing the Zr-Zr distribution (in purple) and the Cr-Zr distribution (in blue). Dashed lines indicate $d_{(0002)}$ spacings from XRD data.}
\end{figure}

Experimental evidence of the dual nature of Cr accommodation in Zr is provided by our APT work. The dataset of the sample is shown in Figure~\ref{fig:APT}(a). A $10 \times 10$ \si{nm^2} subset of the data centred on the $(0002)$ pole indicated in Figure~\ref{fig:APT}(b), and going down the whole length of the dataset, was exported\footnote{Due to the small size of the region, a large portion of those atoms reside at the boundary and have a limited number of neighbours. Therefore, the results taken do not bare a quantitative weight, but can provide qualitative information.}. Advanced species-specific spatial distribution maps were applied to this subset and the resulting data are plotted in Figure~\ref{fig:proxigram}. The Zr-Zr distribution exhibits broad peaks corresponding to the $(0002)$ atomic planes. The slight and progressive shift away from the expected location of the peaks can be attributed to distortions in the tomogram, as discussed in \cite{Gault2011a}. 
The Zr-Cr distribution, which measures the average distribution of Cr atoms relative to Zr atoms along this crystallographic direction, exhibits peaks that are in-between the main peaks of the Zr-Zr distribution. This is consistent with a significant fraction of the Cr atoms being located at interstitial sites. This is the first evidence of interstitials provided by APT. 
A similar procedure was attempted on the Fe-containing sample, but the large volume fraction of SPPs in the sample \cite{Gault2013} made it extremely challenging to achieve a sufficient signal-to-noise ratio to generate definitive results.


\subsection{Lattice expansion of binary alloys}						\label{sec:interstitials}

\begin{figure*}[p]
\centering
	\includegraphics[width=0.75\textwidth]{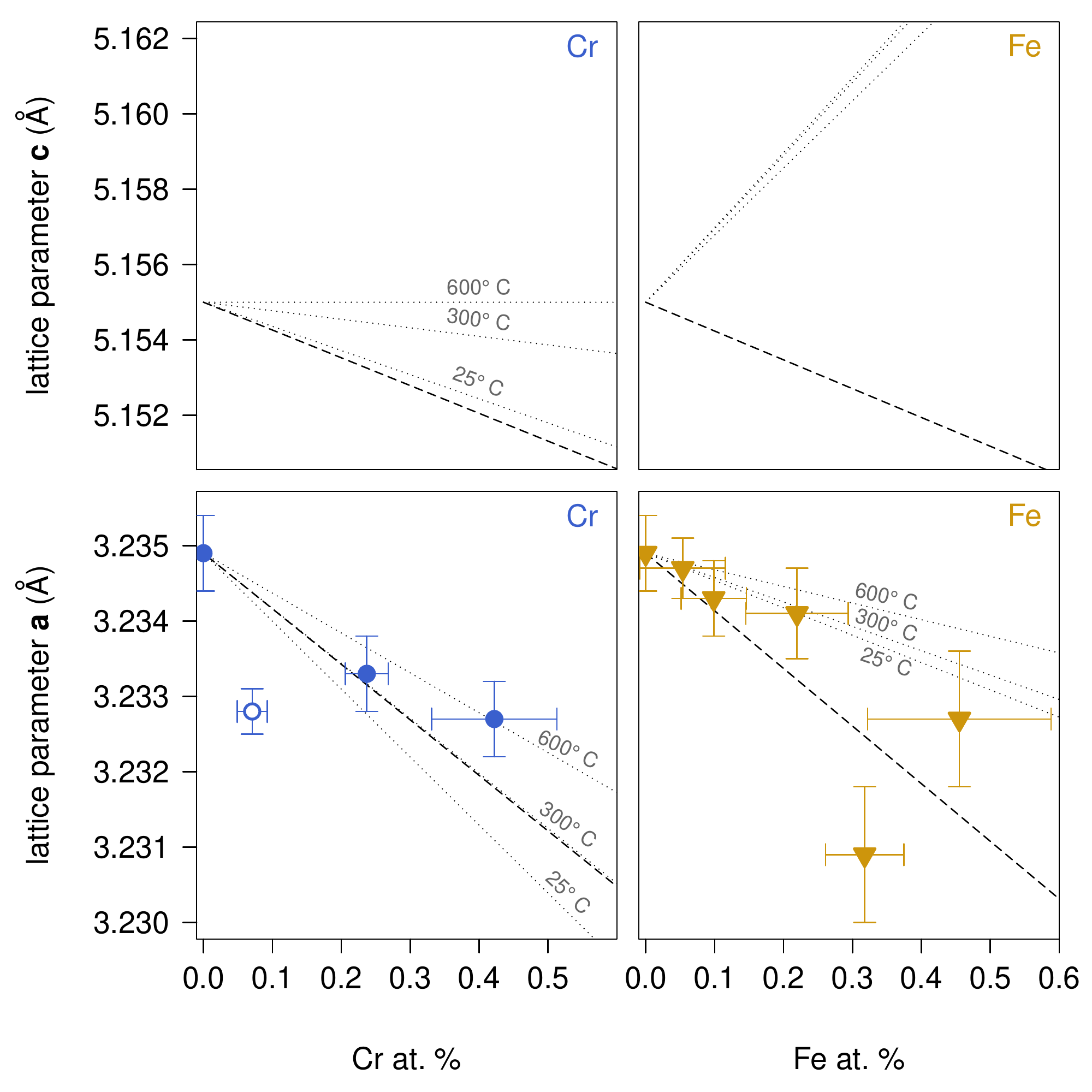}
\caption{\label{fig:expansion} Measured lattice parameter $a$ as a function of alloying element in solution (points) and lattice parameters $a$ and $c$ from theory (lines). Dashed line represents Vegard's law; Dotted lines represent the current prediction of lattice expansion using configurational averaging of DFT simulations of dilute defects. Hollow symbol represent low purity sample. Vertical error bars represent the uncertainty due to machine error, background noise and Rietveld analysis; horizontal error bars represent the compound uncertainty of TEP measurements, APT concentration measurements and standard deviation from linear regression of Figure~\ref{fig:TEP}.}
\end{figure*}

Accommodation of Fe and Cr point defects in the $\alpha$-Zr matrix is predicted to cause noticeable lattice strain (see Table~\ref{tab:interst}). Therefore,
the lattice parameters of the $\alpha$-Zr matrix were measured for all binary samples using XRD. The change in lattice parameter $a$ as a function of alloy concentration is shown in Figure~\ref{fig:expansion}, together with a projection of Vegard's law calculated from the lattice parameters of the pure elements (dashed line) and the DFT predictions of lattice expansion due to the presence of defects (dotted lines). Note that the change in composition and the change in lattice parameter are very small, near the accuracy limit of the XRD equipment, as highlighted by the large error bars.

The DFT predictions rely on configurational averages of the dilute point defects at temperatures of \SI{25}{K}, \SI{300}{K} and \SI{600}{K}. This averaging technique does not include other temperature effects such as thermal expansion or phonon scattering. In other words, the model represents a solution that has been homogenised at those temperatures and subsequently quenched.

Figure~\ref{fig:expansion} shows that, for Cr-Zr solution, both Vegard's law and the DFT predictions are in good agreement with experimental observations, if the low purity sample (hollow circle) is disregarded.  However, for the Fe-Zr solution DFT predictions differ greatly to Vegard's law.
Further, for the $a$ lattice parameter, DFT predictions are in better agreement with experimental data at low concentrations (near the solid solubility of Fe in Zr) but the agreement is somewhat lost at higher concentrations. For the $c$ lattice parameter, DFT results are in stark contrast to Vegard's law in that an expansion is expected instead of a contraction. 
XRD data for the $c$ lattice parameter were inconclusive as the low multiplicity of the $c$ direction caused too much scatter in the data. 
The predicted deviation from Vegard's law suggests that lattice parameter measurements are not a suitable means to estimate the solute concentration of alloying elements.

\begin{figure}[hbt]
\centering
\includegraphics[width=0.49\textwidth]{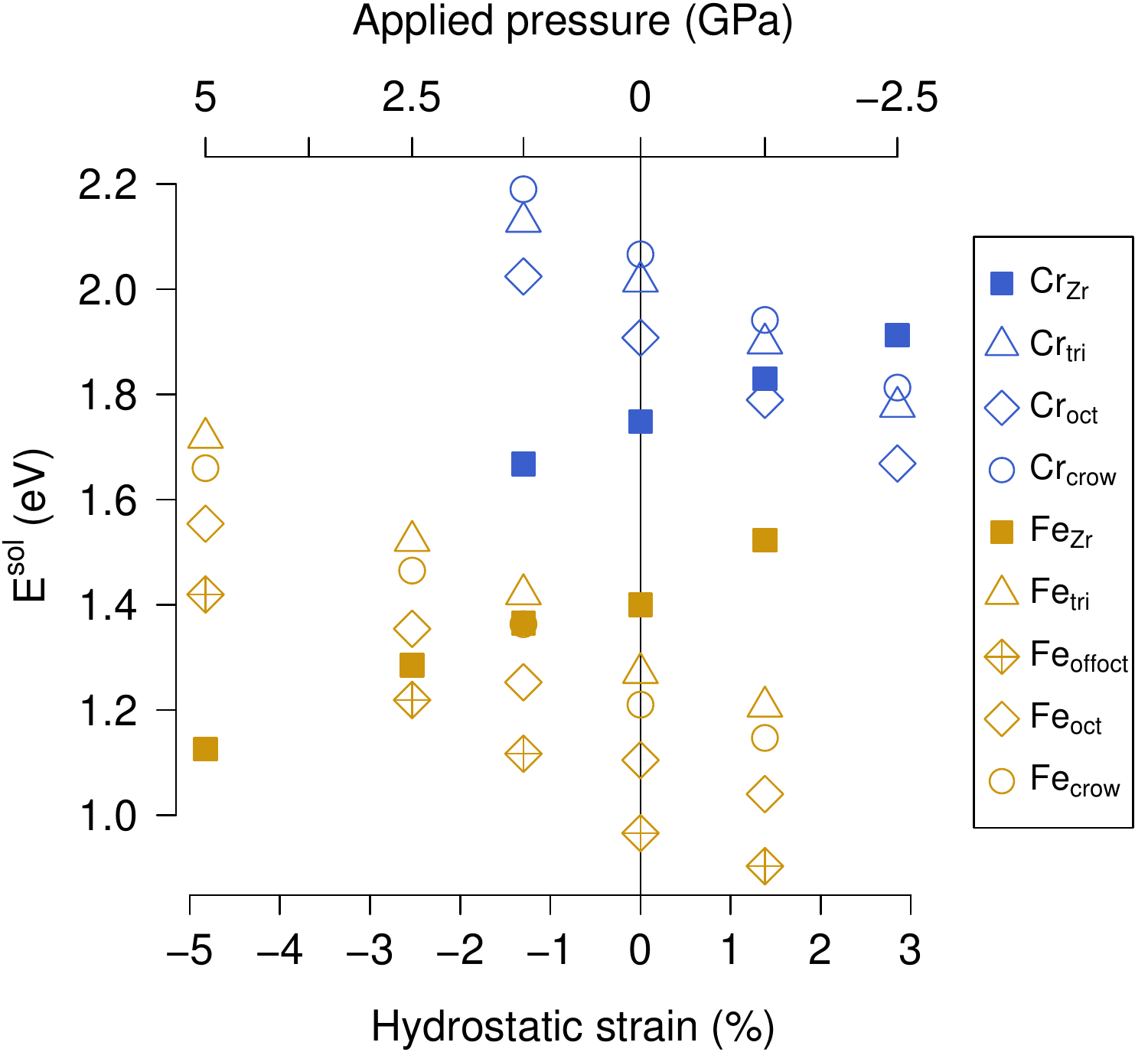}
\caption{\label{fig:press_defect} Energy of solution of Fe (beige) and Cr (blue) accommodated as interstitial species (hollow symbols) and substitution species (filled squares) as a function of hydrostatic strain. The simulation cells were strained prior to adding the defect, by applying an external hydrostatic pressure, displayed in the secondary $x$-axis above (positive = compressive).}
\end{figure}

When performing the configurational average, it is implicit that the defects are not interacting; therefore, strictly, the average is only valid at the dilute limit.
Since the binary solid solutions investigated here are above their respective solubility limits,
it is pertinent to assume that the alloying atoms are interacting with each other. More specifically, the compressive strain field of an interstitial defect is likely to increase the formation energy of another interstitial defect in its vicinity, whilst reducing that of a substitutional defect (which, due to its negative relaxation volume, has a tensile strain field associated with it).
This hypothesis was corroborated by repeating the defect simulations in pre-strained supercells, see Figure~\ref{fig:press_defect}. 
Under a compressive strain, the stability of substitutional defects (filled squares) increases while that of interstitial defects (hollow symbols) decreases; and the opposite is true under a tensile strain.
This helps explain the lack of preference between interstitial and substitutional accommodation that is observed in the APT spatial distribution map of Cr (Figure~\ref{fig:proxigram}).

As well as affecting the relative solution energies, the strain fields of the point defects may provide a driving force for diffusion: defects with opposing strain fields may attract each-other at distances of up to a few angstroms, whilst defects with same-sign strain fields will repel one another. When combined with the extreme mobilities of Fe and Cr \cite{Hood1972,Hood1974,Pasianot2009}, this may lead to the formation of defect clusters with a reduced overall lattice strain, hence a reduced lattice expansion and more favourable solution energy.

\subsection{Cluster formation}								\label{sec:clusters}
\begin{figure*}[hbt]
\centering
\begin{subfigure}[b]{0.249\textwidth}
	\includegraphics[width = \textwidth]{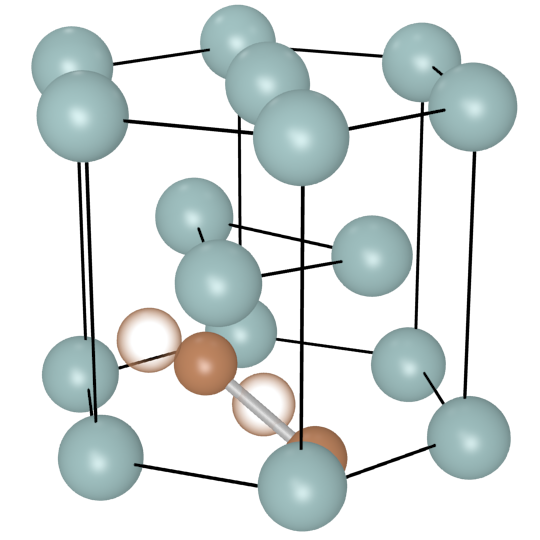}
	\caption{Fe-Fe dumbbell}
\end{subfigure}
\begin{subfigure}[b]{0.249\textwidth}
	\includegraphics[width = \textwidth]{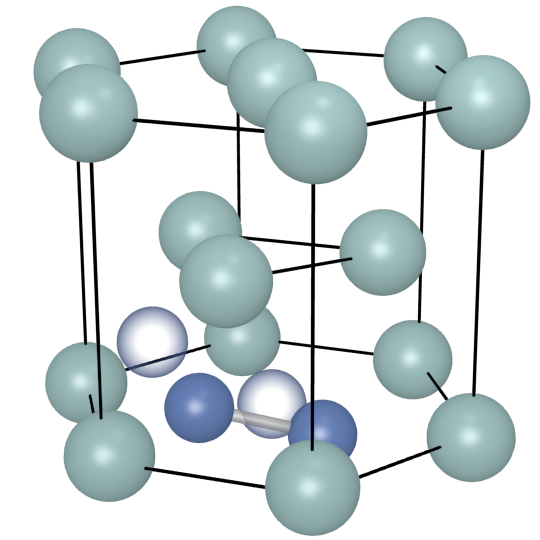}
	\caption{Cr-Cr dumbbell}
\end{subfigure}
\begin{subfigure}[b]{0.249\textwidth}
	\includegraphics[width = \textwidth]{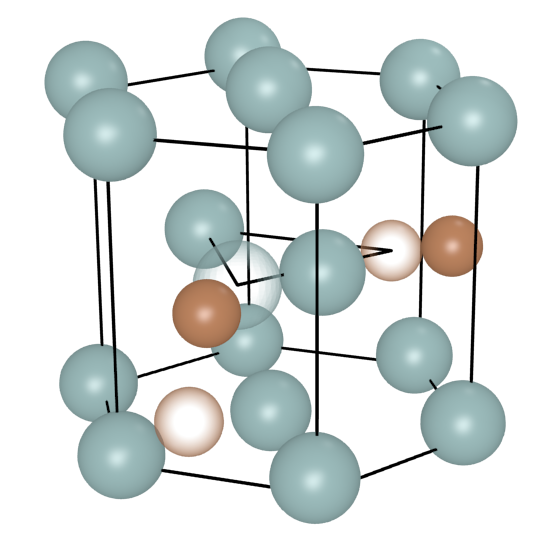}
	\caption{Fe-Fe 4nn}
\end{subfigure}
\begin{subfigure}[b]{0.249\textwidth}
	\includegraphics[width = \textwidth]{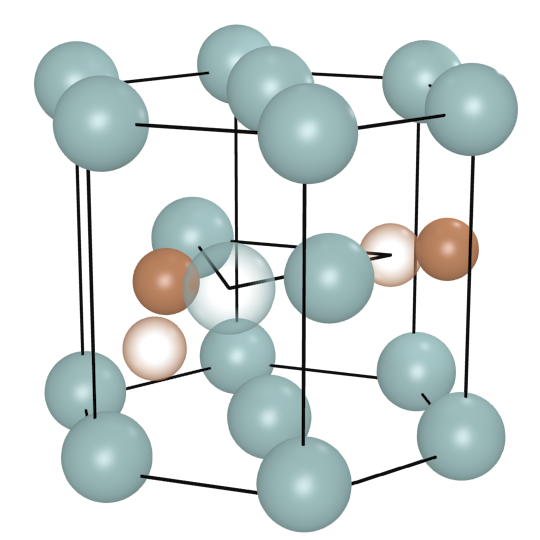}
	\caption{Fe-Fe 3nn}
\end{subfigure}
\begin{subfigure}[b]{0.249\textwidth}
	\includegraphics[width = \textwidth]{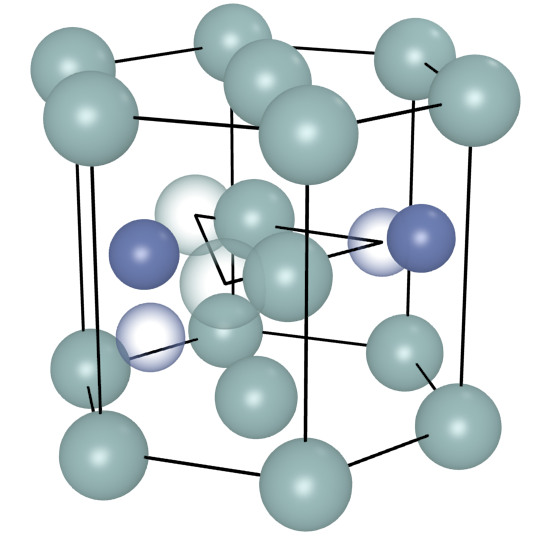}
	\caption{Cr-Cr 3nn}
\end{subfigure}
\begin{subfigure}[b]{0.249\textwidth}
	\includegraphics[width = \textwidth]{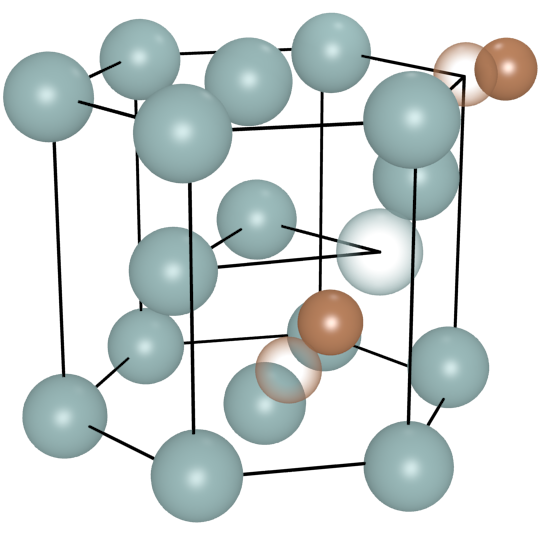}
	\caption{Fe-Fe 5nn}
\end{subfigure}
\caption{\label{fig:clusterConfig} Brown spheres represent Fe atoms, dark blue spheres represent Cr atoms, turquoise spheres represent Zr atoms, translucent spheres represent the initial position of selected atoms.}
\end{figure*}

To investigate the formation of Fe and Cr clusters, simulations containing two extrinsic species were first considered.
The starting positions for the clusters were defined by combining a substitutional defect ($\text{M}_{\text{Zr}}$) with and an octahedral or off-site octahedral interstitial defect ($\text{M}_{i(\text{oct})}$), since these are the most stable defects with opposing strain fields for Cr and Fe respectively (from Section~\ref{sec:interstitials}).
All such configurations that could fit in a $5 \times 5 \times 3$ supercell of $\alpha$-Zr (150 Zr atoms) were investigated, leading to defect-defect separations that range from \SI{2.30}{\angstrom} for the first nearest neighbour (1nn) to \SI{6.97}{\angstrom} for the 7nn configuration.
When considering mixed Fe-Cr clusters, the Cr atoms were placed in the substitutional sites and the Fe atoms in the interstitial sites, $\{\text{Cr}_{\text{Zr}}:\text{Fe}_{i(\text{off-oct})}\}$, owing to the smaller atomic radius of Fe and its preference for interstitial sites (see section~\ref{sec:interstitials}).
The simulations of the clusters were relaxed to a high level of force convergence (\SI{0.05}{eV \angstrom$^{-1}$}) and the atomic positions were perturbed by small amounts in random directions. Furthermore, to provide greater degrees of freedom to the simulations, these were repeated under $\sigma=0$ conditions as well as $\varepsilon=0$ conditions.
This combination ensures that the BFGS minimiser \cite{Byrd1994,Pfrommer1997} is unlikely to trap atoms in local energy minima. In other words, the starting positions are just that, and the extrinsic atoms were expected to explore the energy surface until lowest energy configurations were found.
Figure~\ref{fig:clusterConfig} shows the initial and final configurations of some clusters that have moved from their original lattice sites. In all cases, the resulting defect is an elongated or extended defect, often involving one or more Zr SIAs. Most notably the 1nn clusters moved into a split substitutional (or dumbbell) around a Zr lattice site.  An in-depth analysis of the dumbbell configurations, including a comparison with the intrinsic Zr dumbbells from previous work \cite{Verite2013,Samolyuk2014}, is presented in \ref{sec:dumbbells}.

A summary of the formation energies, binding energies and relaxation volumes of all the 2-atom defects --- in their final relaxed positions --- are presented in Table~\ref{tab:clusters}.
In all cases the dumbbell defect is consistently the most stable configuration, independent of the species involved, and the relative preference for the dumbbell is as high as 0.5--0.6 eV compared to the next most stable configuration.
Importantly, all configurations up to the $5^{th}$ nn are a single lattice jump away from the dumbbell configuration. Whilst this provides incomplete information about kinetics of cluster formation, it does imply that multiple paths exist for migrating extrinsic species to reach (and be trapped in) the dumbbell configuration.

\begin{table}
\scriptsize
\centering
\caption{\label{tab:clusters} Normalised defect formation energy ($E_f$/atom), binding energies ($E_b$), spin and relaxation volumes ($\Omega$), for all defect clusters investigated. The spin is the cumulative spin on the extrinsic elements calculated using Mulliken analysis \cite{Mulliken1955a}. c$[hkjl]$ stands for a crowdion defect along the $[hkjl]$ direction. Nomenclature of dumbbells is defined in \cite{Verite2013} and in \ref{sec:dumbbells}.}
\begin{tabular}{l l S S S S}
\toprule	
		&cluster type		&\text{spin ($\hbar$)}	&\text{$E_{f}$/atom (eV)}	&\text{$E_{b}$ (eV)}	&\text{$\Omega$ (\si{\angstrom^3})}	\\
\midrule
\multicolumn{2}{l}{\bf Fe-Fe}	\\										
1nn	&	\text{dumbbell (PS')} 			&	0.00		&	0.32		&	1.69		&	-2.49	\\
2nn	&	$\{\text{Fe}_{\text{Zr}}:\text{Fe}_{i}\}$	&	1.50		&	0.79		&	0.75		&	1.00	\\
3nn	&	\text{c$[10\bar10]$}				&	0.00		&	0.62		&	1.08		&	-2.39	\\
4nn	&	\text{c$[11\bar20]$}				&	0.00		&	0.57		&	1.18		&	-3.59	\\
5nn	&	\text{c$[20\bar21]$}				&	0.70		&	0.72		&	0.88		&	-1.95	\\
6nn	&	$\{\text{Fe}_{\text{Zr}}:\text{Fe}_{i}\}$	&	1.64		&	0.93		&	0.47		&	2.32	\\
7nn	&	$\{\text{Fe}_{\text{Zr}}:\text{Fe}_{i}\}$	&	1.65		&	0.96		&	0.42		&	2.00	\\
\midrule											
\multicolumn{2}{l}{\bf Cr-Cr}	\\										
1nn	&	\text{dumbbell (P2S)} 			&	0.00		&	0.95		&	1.95		&	1.92	\\
2nn	&	$\{\text{Cr}_{\text{Zr}}:\text{Cr}_{i}\}$	&	2.30		&	1.26		&	1.34		&	7.13	\\
3nn	&	\text{c$[10\bar10]$}				&	0.00		&	1.41		&	1.03		&	1.81	\\
4nn	&	\text{c$[11\bar20]$}				&	2.26		&	1.31		&	1.22		&	5.52	\\
5nn	&	\text{c$[20\bar21]$}				&	0.00		&	1.54		&	0.76		&	2.07 \\
6nn	&	$\{\text{Cr}_{\text{Zr}}:\text{Cr}_{i}\}$	&	2.29		&	1.37		&	1.11		&	2.96	\\
7nn	&	$\{\text{Cr}_{\text{Zr}}:\text{Cr}_{i}\}$	&	2.29		&	1.36		&	1.11		&	7.91	\\
\midrule											
\multicolumn{2}{l}{\bf Fe-Cr}	\\										
1nn	&	\text{dumbbell (PS')} 			&	0.00		&	0.64		&	1.77		&	-0.42	\\
2nn	&	$\{\text{Cr}_{\text{Zr}}:\text{Cr}_{i}\}$ 	&	0.00		&	1.17		&	0.70		&	0.66	\\
3nn	&	$\{\text{Cr}_{\text{Zr}}:\text{Cr}_{i}\}$ 	&	2.32		&	0.89		&	1.26		&	6.89	\\
4nn	&	\text{c$[11\bar20]$}				&	2.14		&	0.87		&	1.31		&	-0.20	 \\
5nn	&	\text{c$[20\bar21]$}				&	2.19		&	0.76		&	1.52		&	5.40	\\
6nn	&	$\{\text{Cr}_{\text{Zr}}:\text{Cr}_{i}\}$	&	0.00		&	1.36		&	0.32		&	2.96	\\
7nn	&	$\{\text{Cr}_{\text{Zr}}:\text{Cr}_{i}\}$	&	2.27		&	0.90		&	1.25		&	6.34	\\
\bottomrule
\end{tabular}
\end{table}

With regard to the lattice expansion, all two-atom clusters exhibit relaxation volumes that are significantly smaller than those of the single-atom dilute defects.
Furthermore, the combined defect volume of dilute Fe$_\text{Zr}$ and Fe$_i$ is \SI{3.13}{\angstrom^3}, while that of the bound dumbbell is only \SI{-2.49}{\angstrom^3}.
Similarly for Cr, the combined volume of dilute defects is \SI{3.81}{\angstrom^3} compared with \SI{-1.95}{\angstrom^3} for the dumbbell. Finally, in the mixed case (in which Cr takes the substitutional site and Fe takes the off-oct interstitial site), the combined volume of dilute defects is again greater than that of the dumbbell (\SI{2.50}{\angstrom^3} against \SI{-0.42}{\angstrom^3}).
This suggests that part of the binding energy comes from a reduction of lattice strain.

Notably, many of the Fe-Fe defect pairs exhibit negative relaxation volumes, resulting in a tensile strain field, despite the addition of one extra atom in the supercell. In particular, the most favourable configurations (1nn, 3nn \& 4nn) exhibit tensile strain fields arising from relaxation volumes of \SI{-2.49}{\angstrom^3}, \SI{-2.39}{\angstrom^3} and \SI{-3.59}{\angstrom^3}.

\begin{figure}[hbt]
\centering
\begin{subfigure}[b]{0.22\textwidth}
	\includegraphics[width = \textwidth]{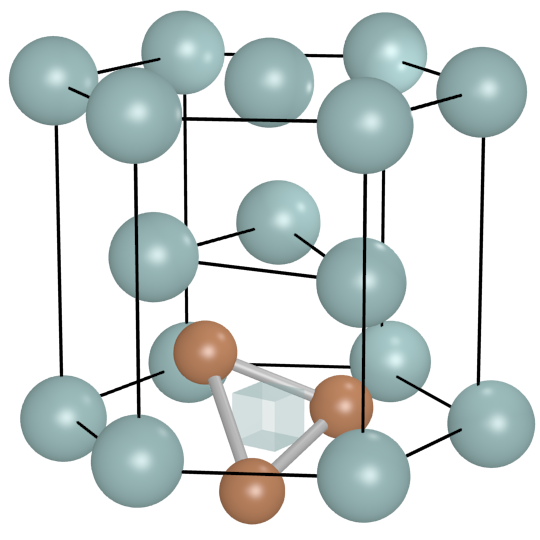}
	\caption{\{3M\}$_\text{Zr}$}
\end{subfigure}
\begin{subfigure}[b]{0.22\textwidth}
	\includegraphics[width = \textwidth]{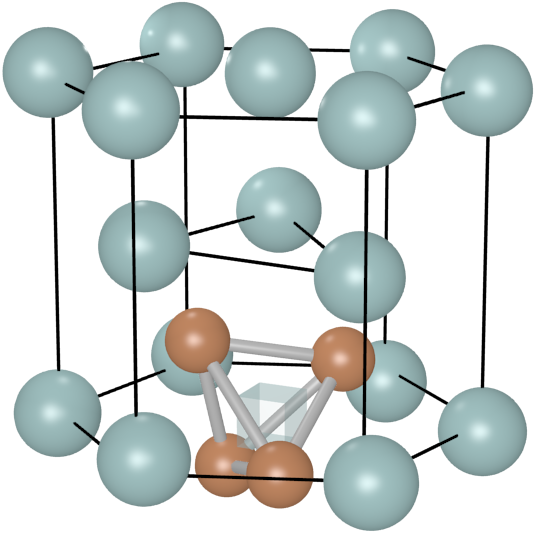}
	\caption{\{4M\}$_\text{Zr}$}
\end{subfigure}
\caption{\label{fig:3-4Fe} Brown spheres represent Fe or Cr atoms, turquoise spheres represent Zr atoms and the translucent cube represents the Zr vacancy.}
\end{figure}

\begin{figure}[hbt]
\centering
\includegraphics[width=0.49\textwidth]{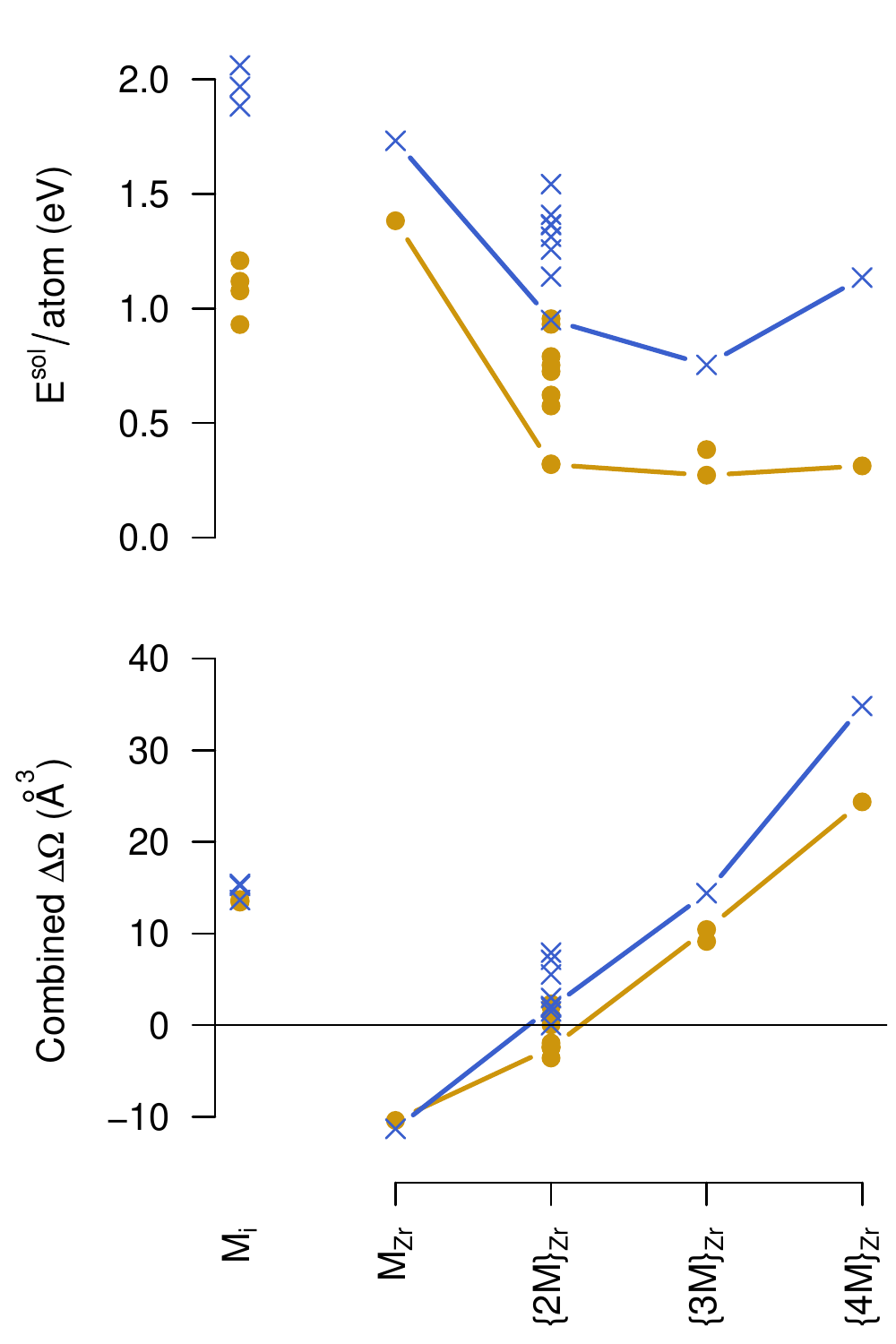}
\caption{\label{fig:clusterSol} Solution energy (top) and relaxation volume (bottom) as a function of cluster size. Dilute interstitial defects are also included for comparison. Joined points show the lowest energy clusters. Blue crosses represent Cr defects, yellow dots represent Fe defects.}
\end{figure}

To investigate if more than two Fe or Cr atoms could be accommodated in or around a single Zr vacancy, a third and then a fourth interstitial atom were added to the relaxed dumbbell configurations, as these are the most stable 2-atom clusters (see Figure~\ref{fig:3-4Fe}).
The resulting solution energies (normalised per extrinsic atom, as defined in \cite{Burr2013}) and defect volumes are presented in Figure~\ref{fig:clusterSol}.
Clusters containing 3 and 4 Fe atoms exhibit similar solution energies to the 2-atom clusters. Furthermore, from Figure~\ref{fig:clusterSol} (top) it is clear that even the least stable 2-atom clusters exhibit solution energies that are similar to the most stable dilute defect.
The relaxation volumes of the clusters containing a Zr vacancy shows a fairly linear behaviour, see Figure~\ref{fig:clusterSol} (bottom). As a result, clusters containing 3 Fe or Cr atoms cause a smaller expansion of the lattice than a dilute interstitial Fe or Cr defect.

\begin{figure}[hbt]
\centering
\includegraphics[width=0.49\textwidth]{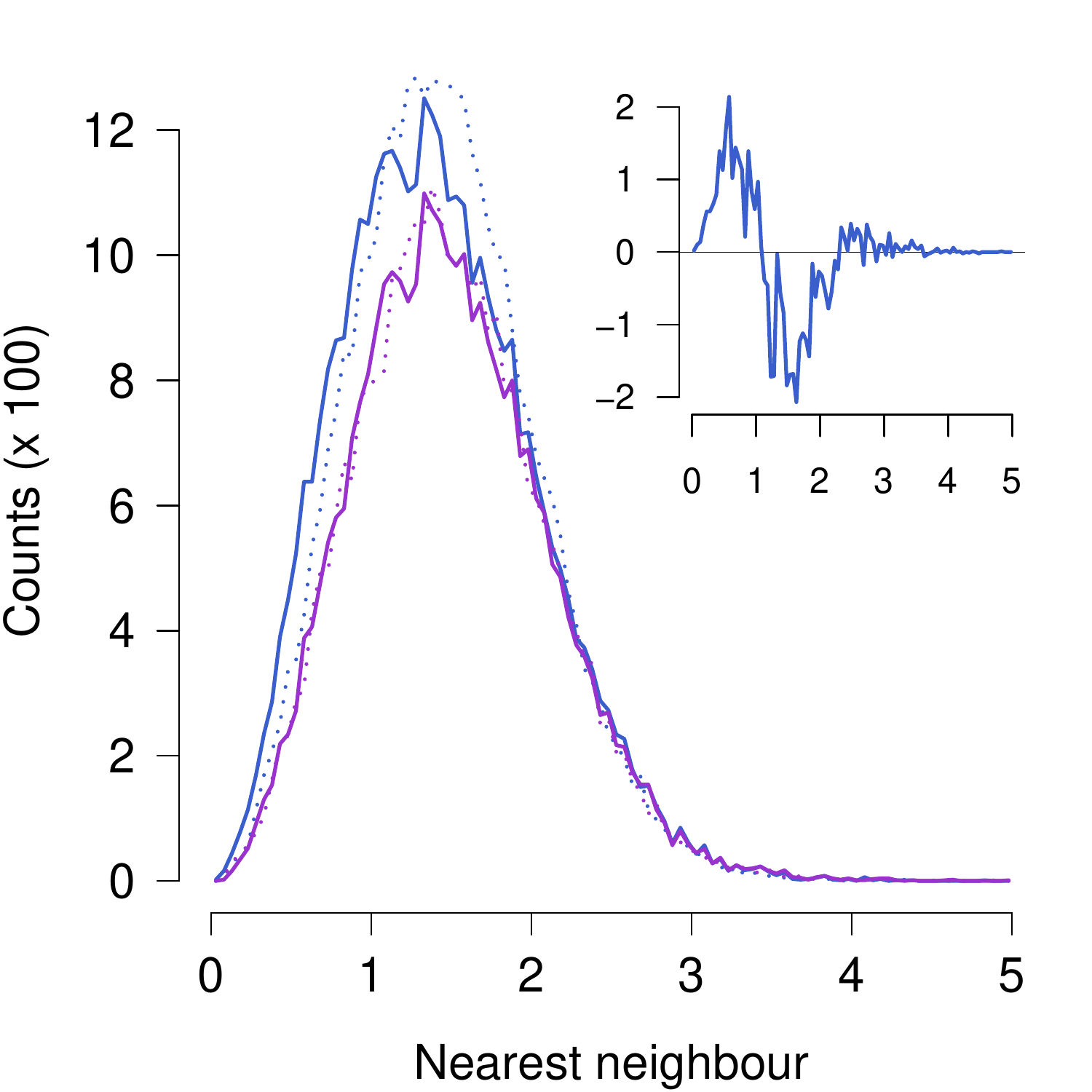}
\caption{\label{fig:NN_distro} Cr first nearest neighbour distribution (solid lines) compared to randomised datasets (dotted lines) for the whole dataset (blue) and for the matrix only (purple). Inset is the difference between the experimental and random that highlights the clustering tendency of Cr in the full dataset.}
\end{figure}

Clustering behaviour of Cr was also investigated using APT.
Two distinct analyses were performed to investigate the distribution of distance between each Cr atom and their first nearest neighbour Cr: first considering the whole dataset, and second considering a subset of the data that excludes the regions delineated by an isoconcentration surface similar to the one displayed in Figure~\ref{fig:APT}, with a threshold of \SI{1}{at\% Cr}. The latter allows for an analysis of the matrix. The two graphs, shown in Figure~\ref{fig:NN_distro}, exhibit a different behaviour: for the complete dataset, there is a clear tendency for neighbours of shorter distances compared to random, while Cr in the matrix are close to a random distribution, which was expected based on visual inspection. The closer distances in the subset containing the clusters, is in agreement with the DFT predictions that closely bound clusters (those in 1nn configuration) are more stable than clusters in which the Cr atoms are further apart and compared to dilute defects.

Interestingly, and as shown in Figure~\ref{fig:APT}, the Cr-rich regions are seen to align along specific directions, which could be twin or low-angle boundaries (a high-angle grain boundary would have been identifiable in the dataset via a change in the pole structure \cite{Gault2012}), or dislocation loops, which are likely to be defective areas within the material. This could imply that for Cr to cluster there needs to be some defects in their vicinity.

\subsection{Implication for re-solution of Fe and Cr in irradiated Zr}
A marked increase in corrosion rates of Zr cladding is observed at high fuel burn-up \cite{Garzarolli1996,Griffiths2011}. Due to the delay of the irradiation effect, the increase in corrosion rate is attributed to the amorphisation and dissolution of SPPs that also occurs at high burnup \cite{Garzarolli1996,Griffiths2011}.
Here we consider the fate of Fe and Cr,  that are released into the $\alpha$-Zr matrix following SPP dissolution.

DFT and APT results provide strong indications that in the presence of a Zr vacancy, the formation of clusters significantly lowers the solution energy of Fe and Cr in $\alpha$-Zr, thereby increasing their apparent solubility in $\alpha$-Zr.
The extremely favourable binding energies observed for all the two-atom clusters suggest that if a cluster is formed, there will be a barrier against separation. If the clusters do not separate, the only other fate is growth. It is well known that Fe and Cr are ultra-fast diffusers in $\alpha$-Zr \cite{Hood1974,Tendler1987,Pasianot2009}, it is, therefore, likely that a third atom will fall within the proximity of the cluster and also get trapped. Binding energies were found to remain strongly positive with the addition of third or fourth Fe atoms. Whilst the simulation of larger clusters is unfeasible due to computational limitations, it is reasonable to expect that larger clusters may be favourable if more that one Zr vacancy was present (e.g.\ considering the interaction of Fe or Cr with the vacancy clusters investigated by \cite{Varvenne2014}, or with gliding dislocations \cite{Voskoboinikov2005a,Voskoboynikov2005}).

Since the mobility of Fe and Cr is not likely to be a limiting step in cluster formation, the most probable limiting factor is the presence of Zr vacancies.
Fe in solution in Zr naturally takes an interstitial position (see section~\ref{sec:interstitials}), 
whilst all the clusters considered in the current work contain one Zr vacancy. Therefore, 
for the formation of an Fe/Cr cluster, an additional Zr vacancy must be present as well as at least two interstitial atoms.
Further confirmation of this comes from APT, which suggests the need for defects nearby, for the formation of Cr clusters.


Under equilibrium conditions the concentration of Zr vacancies is very low (our DFT calculated vacancy formation energy of \SI{1.9}{eV} yields a concentration of $3.7\times10^{-9}$ at the $\beta \rightarrow \alpha$ transition temperature), and therefore only a limited number of defect clusters may form. On the other hand, upon irradiation the concentration of vacancies and dislocation loops increases by orders of magnitude  \cite{Griffiths1988}.
Neutron irradiation also causes SPPs to become amorphous and to leach Fe and Cr back into the $\alpha$-Zr matrix \cite{Yang1986,Yang1988,Griffiths1987,Lefebvre1990,Griffiths1990,Griffiths1995,Kakiuchi2006,Cockeram2014}. In view of the current results, this re-solution can be explained by the presence of additional Zr vacancies introduced through irradiation, which promotes the formation of defect clusters.
From this point of view, cluster formation may be considered as a competing mechanism to SPP formation.
The classical (macroscopic) notion of a single $\alpha$-Zr phase with a uniform random distribution of alloying additions is not representative of the irradiated Zr-Fe-Cr system at an atomic scale.
A better description is one consisting of nano-sized clusters containing most of the dissolved Fe and Cr, separated by volumes of near pure $\alpha$-Zr.

\section{Summary}
Binary Fe-Zr and Cr-Zr alloys were investigated using a combination of experimental and modelling techniques.
Buttons of binary alloys were manufactured with varying concentrations of alloying elements and then quenched at fast cooling rates in an attempt to hinder nucleation of SPPs. TEM investigation shows that SPPs had formed in all samples. Nonetheless, TEP measurements revealed that an increasing amount of the alloying elements were trapped into solution with increasing nominal composition of the samples. APT was employed to measure the solution concentration of one Fe-Zr sample and one Cr-Zr sample. A linear regression following the trend of the TEP results was used to estimate the solution concentration in the other binary samples.

Results from APT demonstrated that Cr can occupy interstitial sites, and that Cr tends to cluster along linear features, which are likely to be related to defective regions of the sample.
In turn, this provides confidence in the DFT results, which predicted very close energies for the accommodation of Cr as an isolated interstitial or substitutional species, as well as that Cr tends to cluster near vacancies. On the other hand, Fe atoms in dilute conditions exhibit a distinct preference for only interstitial accommodation. Furthermore,
 two low-symmetry interstices (the crowdion and the off-site octahedral) were successfully modelled for the first time. The off-site octahedral was found to be the most favourable site for the accommodation of Fe, whilst the tetrahedral site, which has previously been modelled, was found to be unstable.

XRD results show that limited expansion or contraction of the $a$ lattice parameter is observed with increasing Fe or Cr content. Vegard's law provides a reasonable approximation to the change in lattice parameter $a$ for Cr. However, for Fe-Zr solutions, a deviation from Vegard's law is observed in agreement with DFT simulations. The same calculations also indicate a largely different behaviour for the $c$ lattice parameter, which is predicted to expand whilst Vegard's law predicts a contraction. DFT predictions based on single point defects fall short at higher Fe concentrations, where the dilute non-interacting assumption is no longer valid.
At those higher concentrations, the formation of small clusters composed of defects with opposing strain fields is predicted.

DFT calculations show that 2-atom clusters exhibit much smaller defect volumes compared to their dilute defect components.
A variety of different clusters were identified which were strongly bound and more stable than their dilute counterparts, thereby reducing the solution energy of Fe and Cr in $\alpha$-Zr. In particular, dumbbell configurations (i.e.\ two extrinsic elements around a Zr vacancy) were most favourable. The nearest neighbour distribution from atom probe microscopy confirms a prevalence of dumbbell configurations for Cr clusters.
The formation of clusters is predicted to be limited by the presence of intrinsic Zr vacancies.
This helps explain the re-solution of Fe and Cr in $\alpha$-Zr, which is observed upon neutron irradiation, as a direct consequence of a higher intrinsic defect population.
The current work predicts that up to four Fe atoms or three Cr atoms can be accommodated around a single Zr vacancy. Larger clusters involving more than one Zr vacancy --- which are currently computationally impractical to model --- are likely to form in the vicinity of strain gradients (caused by grain boundaries, dislocations, voids, etc.) and could further reduce the solution energies of Fe and Cr.

\section{Acknowledgements}
Computing resources were provided by the Imperial College HPC, UK and the MASSIVE supercomputer facility, Melbourne, Australia.
P.~A.~Burr acknowledges ANSTO and the EPSRC for financial support.
B.~Gault acknowledges funding as well as scientific and technical input from the Australian Microscopy \& Microanalysis Research Facility (AMMRF) at the University of Sydney. B.~Gault also acknowledges that he is a full time employee of Elsevier Ltd.\ but declares no conflict of interest as this work was performed in his spare time. M.~Ivermark acknowledges Westinghouse Electric Sweden for funding.
Western Zirconium, Utah, USA for provision of material and INSA, Lyon, France for TEP resources, are also acknowledged.

\appendix{}
\section{Relaxation tensors of point defects}
\label{App:relaxTensor}
The relaxation volume tensors of the dilute point defects from DFT simulations were calculated using the \emph{aneto} program and are presented in Table~\ref{tab:stressTensors}. The values are taken from $\varepsilon=0$ simulations.
\begin{table*}[hbt]
\small
\centering
\caption{\label{tab:stressTensors} Components of the relaxation volume tensor in units of \si{\angstrom^3}.}
\begin{tabular}{l l S S S S S S}
\toprule
&	&\text{$\Omega_{11}$}	&\text{$\Omega_{22}$}	&\text{$\Omega_{33}$}	&\text{$\Omega_{12}$}		&\text{$\Omega_{13}$} &\text{$\Omega_{23}$} \\	
\midrule
Fe	&off-site substitution		&-2.15	&-4.12	&-4.53	&1.71	&0.18	&-0.32	\\
	&substitution			&-3.94	&-3.94	&-8.36	&0.00	&0.00	&0.00	\\
 	&octahedral			&4.69	&2.87	&6.42	&1.58	&-0.08	&0.14	\\
	&off-site octahedral 		&4.03	&3.84	&5.95	&2.75	&-0.51	&-0.79	\\
	&trigonal				&5.76	&6.67	&-0.18	&3.20	&0.00	&0.00	\\
	&crowdion				&5.14	&4.52	&3.86	&1.69	&1.71	&2.92	\\
\midrule
Cr	&off-site substitution		&-1.90	&-3.85	&-5.57	&-0.16	&0.00	&0.00	\\
	&substitution			&-4.00	&-4.00	&-4.83	&0.00	&0.00	&0.00	\\
 	&octahedral			&5.03	&3.30	&6.87	&1.50	&-0.05	&0.09	\\
	&trigonal				&6.14	&6.35	&1.16	&2.99	&0.01	&-0.01	\\
	&crowdion				&5.62	&5.06	&4.73	&0.48	&1.77	&-3.06	\\
\bottomrule
\end{tabular}
\end{table*}

\section{Dumbbell configurations}
\label{sec:dumbbells}
Here we employ the nomenclature introduced by V\'erit\'e \etal~\cite{Verite2013}, to refer to the orientation of the dumbbell with respect to the Zr lattice:  vertical (S), basal (BS), rotated on prismatic plane type I (PS or PS' depending on the rotation angle) and rotated on prismatic plane type II (P2S). 

The Fe-Fe dumbbell is rotated in the prismatic plane type I $(11\bar20)$ by a tilt angle of \SI{54.5}{\degree} from the \emph{c}-axis, in accordance with the PS' defect observed in Zr SIAs. The extrinsic dumbbell exhibited lower symmetry than the intrinsic SIAs: the Fe dumbbell is $\sim0.25$\si{\angstrom} out of plane with respect to the $(11\bar20)$ plane, and also off-centre with respect to the host Zr vacancy so that the one Fe atom is closer to the vacancy than the other (the centre of the dumbbell is displaced by $\sim0.005$\si{\angstrom} with respect to the host Zr vacancy).
These simulations were repeated up to five times with different starting configurations and tight force convergence criteria to ensure that such subtle measurements were not artefacts caused by computational parameters. 
One other stable position was observed for the Fe dumbbell when rotated so that it would align vertically. Again the dumbbell is preferentially located $\sim0.40$\si{\angstrom} out of plane. This configuration is 
\SI{0.861}{eV} less stable than the one described above, and was found to lie on a shallow minima near the transition state for rotation of the dumbbell.
Unlike Zr SIAs, no other stable orientations were observed.


The Cr-Cr dumbbell exhibits a configuration similar to the P2S defect (as opposed to the PS' of the Fe-Fe dumbbell), with a large tilt angle of $\theta = 76.4$\si{\degree}. However, the defect is not exactly on the type II prismatic plane $(10\bar10)$, instead it exhibits a small deviation angle of $\psi = 8.2$\si{\degree} from the $(10\bar10)$ plane. Furthermore, the centre of the dumbbell is \SI{0.172}{\angstrom} below the Zr lattice.


The mixed element dumbbell exhibits similar properties to the Fe-Fe dumbbell: it adopts the PS' orientation with a tilt angle of $\theta = 65.9$\si{\degree}, but in this case the dumbbell is in-plane but slightly off-centre (shifted \SI{0.321}{\angstrom} in the \emph{c}-direction).

The difference between the dumbbells investigated here and SIA dumbbells, are thought to be related to the different ionic sizes and electronegativity of the extrinsic elements. Therefore, whilst the differences may appear to be subtle, they are important for the development of accurate larger scale models, such as classical potentials for molecular dynamics, or kinetic Monte Carlo simulations.



\section*{References}

\bibliographystyle{model1a-num-names}

\bibliography{/Users/pab07/Documents/papers/library}			

\clearpage

\end{document}